\documentclass[conference]{IEEEtran}
\IEEEoverridecommandlockouts
\usepackage{cite}
\usepackage{bm}
\usepackage{multicol}
\usepackage{multirow}
\usepackage[cmex10]{amsmath}
\usepackage{graphicx}
\usepackage{url}
\usepackage{color}
\usepackage{amsfonts}
\usepackage{amsmath}
\usepackage{extarrows}
\usepackage{graphicx}
\usepackage{caption}
\usepackage{algorithm}
\usepackage{algpseudocode}
\usepackage{indentfirst}
\usepackage{esint} 
\usepackage{graphics}
\usepackage{epsfig}
\usepackage{soul}
\usepackage{gensymb}
\usepackage{balance}

\usepackage{algorithm,algpseudocode}
\makeatletter
\newcommand\fs@spaceruled{\def\@fs@cfont{\bfseries}\let\@fs@capt\floatc@ruled
  \def\@fs@pre{\vspace{1\baselineskip}\hrule height.8pt depth0pt \kern2pt}%
  \def\@fs@post{\kern2pt\hrule\relax}%
  \def\@fs@mid{\kern2pt\hrule\kern2pt}%
  \let\@fs@iftopcapt\iftrue}
\makeatother
\def\bf{{\mathbf{f}}}

\begin{document}
%
\title{Channel Estimation and Hybrid Architectures for RIS-Assisted Communications}
%
%
%

\author{Jiguang~He, Nhan~Thanh~Nguyen, Rafaela~Schroeder, Visa~Tapio, Joonas~Kokkoniemi, and~Markku~Juntti\\
     Centre for Wireless Communications, FI-90014, University of Oulu, Finland\\}

\maketitle

\begin{abstract}
Reconfigurable intelligent surfaces (RISs) are considered as potential technologies for the upcoming sixth-generation (6G) wireless communication system. Various benefits brought by deploying one or multiple RISs include increased spectrum and energy efficiency, enhanced connectivity, extended communication coverage, reduced complexity at transceivers, and even improved localization accuracy. However, to unleash their full potential, fundamentals related to RISs, ranging from physical-layer (PHY) modelling to RIS phase control, need to be addressed thoroughly. In this paper, we provide an overview of some timely research problems related to the RIS technology, i.e., PHY modelling (including also physics), channel estimation, potential RIS architectures, and RIS phase control (via both model-based and data-driven approaches), along with recent numerical results. We envision that more efforts will be devoted towards intelligent wireless environments, enabled by RISs. 
\end{abstract}

\IEEEpeerreviewmaketitle

\section{Introduction} \label{sec: intro}
The upcoming sixth-generation (6G) communication system is expected to significantly enhance the quality of services (QoSs). To this end, the key performance indicators in the current fifth generation (5G)
need to be further improved. This can be reaped by Terahertz (THz) communications, reconfigurable intelligent surfaces (RISs), and their integration with artificial intelligence (AI)~\cite{huang2019holographic,Letaief-8808168,Wu-9326394}. Among these potential candidates, a RIS, deployed with a large number of low-cost passive elements, is envisioned to provide the communications with a higher spectral efficiency/energy efficiency (SE/EE)~\cite{Wu2020Mag}. A RIS can have various functionalities, e.g., reflection, polarization, refraction, and absorption, making intelligent control of the wireless propagation environment feasible~\cite{huang2018,huang2019holographic,Shlezinger2020}. With the introduction of the RIS to wireless communications, a new era, in which a controllable propagation environment becomes a reality, is deemed to come. Besides for the communications, the RIS can also be considered for localization together with millimeter wave (mmWave) multiple-input multiple-output (MIMO) networks~\cite{wymeersch2019radio,He2019large,he2019adaptive}. 

In terms of hardware implementation, the RIS can be made of varactor diodes, crystal liquid, etc~\cite{4286010-hum,Perez-Palomino-7109872}. As for physical-layer (PHY) modelling, the recent papers~\cite{Najafi2020,Garcia2020,ozdogan2020} focused on the analyses of path loss for different RIS behaviors, i.e., reflection vs.\ scattering, and the RIS phase profile is designed by addressing the corresponding integral equations for the electric and magnetic vector fields. Alternative physically feasible fading models were proposed under isotropic/rich scattering in~\cite{Bjornson2021,alexandropoulos2021reconfigurable}. It should also be noted that a ray-tracing-based RIS channel model can be found in~\cite{basar2020indoor} for both indoor and outdoor environments.

In a RIS-assisted network, the channel state information (CSI) acquisition is essential for the RIS phase control, beamforming, resource allocation, and interference management~\cite{Abeywickrama2020}. Thus, efficient channel estimation (CE) algorithms are necessary. The RIS is often integrated into mmWave or (sub-)THz communications systems to enable line-of-sight (LOS)-like connectivity. In these frequency bands, the wireless channels are usually very directive and sparse with limited numbers of resolvable paths. Thus, the channel coefficient matrices are rank-deficient, resulting in inherent channel sparsity. A variety of compressive sensing (CS) techniques, e.g., atomic norm minimization, basis pursuit, approximate message passing (AMP), and mixed norm minimization, can be well-tailored for CE~\cite{Wang-9103231,he2020channel,he2020channelANM,Ardah2020trice,schroeder2020passive,Wei2021} in the RIS-aided system. In order to collect measurements, pilot sequences or beam training matrices are required. 
In the parametric channel models, heavily used in mmWave and THz communications, the channel parameters and locations of the terminals are closely coupled. Knowing one can infer the other. Thus the prior information on the mobile station (MS) or environment objects will facilitate the design of pilot sequences, which in turn improves the CE performance~\cite{he2020leveraging}. In addition to conventional model-based approaches, data-driven approaches, for instance, deep learning (DL) frameworks, can also be employed for CE, phase control, and symbol detection in RIS-aided communication systems. They are capable of mapping the received signals to the individual channels, RIS phase profile, and constellation points depending on the purpose of data-driven models.

RIS-assisted communication has gained much interest recently. Various researches have been conducted to realize the RIS, including but not limited to CE, performance analysis, design and optimization of the RIS. In this paper, we offer an overview of the recent research progress and results on RIS technology. 
In particular, we at first provide discussion on the physical modelling for the RIS. Then, we introduce efficient CE algorithms, novel architectures, and data-driven methods for the RIS-assised communication system.





\section{Physical Layer Modelling} \label{Sec: Physical_layer_modelling}
\subsection{Physical RIS Modelling}
In the system models used in algorithm development, the RIS is typically modeled as an array of independent reflecting elements separated by $\lambda/2$ ($\lambda=$ wavelength). This assumption is valid if the RIS is implemented as a reflecting antenna array. Indeed, electrically tunable reflecting antenna arrays can be used to dynamically adjust their radiation patterns \cite{4286010-hum}. The control of the antenna elements can be implemented by loading the antenna elements with tunable impedances. The most often used tunable impedance in RIS prototypes is a varactor implemented with a p-i-n diode. By varying the load impedance, the resonance frequency of an antenna element is varied. This leads to the change in the phase response of the element at the frequency of the impinging wave. The desired reflecting pattern of the RIS can then be realized by joint control of the reflection phases of individual elements in a similar manner as in antenna arrays used in wireless receivers and transmitters.

However, in many of the publications discussing the potentials of the RIS in wireless systems, the RIS is assumed to be implemented with a metasurface. Those are electrically thin and dense two-dimensional arrays of structural elements possessing desired properties granted by their constitutive elements. Elements are called meta-cells, unit-cells or meta-atoms. The size of a meta-cell is typically much smaller than the signal’s wavelength, between $\lambda/10$ and $\lambda/5$ \cite{GLYBOVSKI20161,LIASKOS20191}. Metasurfaces have the potential to offer better control on the propagation of the electromagnetic (EM) waves than the reflecting antenna arrays. In addition to steering the reflection or refraction angles of the impinging EM wave, metasurfaces can be designed to, e.g., absorb the waves to reduce the interference, change the polarization of the waves, and realize data modulation \cite{huang2018,9140329-renzo}. 

\subsection{RIS Channel Modelling}

Understanding the behavior of frequency bands is crucial to optimize the frequency resources among different applications. The trend is to use higher frequencies in order to provide large capacities for data-hungry applications. The mmWave and THz band communications offer very high bandwidths with the cost of large propagation loss. Those are handled with highly directional high gain antenna systems. The large losses, especially in the non-line-of-sight (NLOS) paths, ultimately mean that the high-frequency channels tend to be sparse. That is, there are only few utilizable high gain paths and the common assumption of a rich scattering channel is no longer valid. Channel sparsity leads to challenges in aligning the transmit and receive beams and also to weak connections. A RIS may help by modifying the propagation channels by increasing the gain or creating a reflected LOS connection.

The channel modelling with RISs consists of the channel between the base station (BS) and the RIS, the RIS response, and the RIS--MS channels. As such, the BS--RIS and RIS--MS channels require novel channel models for mmWave and THz bands. There are many recent works aiming modelling mmWave and THz band propagation channels, such as \cite{Akdeniz2014,Haneda2016}. The specific RIS channel models depend not only on the frequency, but also on the placement scheme of the RISs. As RISs and BSs are stationary in the general, their positions can be optimized to minimize the channel path losses \cite{Ntontin2020}. The MSs are in general randomly located and mobile. This causes a need to optimize the RIS beamformers in order to maximize the end-to-end link quality. From the channel point of view, the CE depends on the specific frequency-dependent path loss models. However, the mmWave and THz band channels share the sparsity due to weak penetration through the objects as well as lower dynamic range caused by large bandwidths. That is, the weak paths tend to disappear due to the high noise floor. On the other hand, to overcome the large path loss, directional antennas are required also in the CE phase. This is a challenge that we will be looked into below. In summary, the high-frequency channels are very promising due to large available frequency resources, but they also impose challenges in coping with low dynamic range and large path loss. The RISs are seen as very promising technologies to maximize the NLOS channel gains and to make it possible to increase the service area of the BSs and increases the QoSs at the same time.

\section{RIS Channel Estimation}
\begin{figure}
    \centering
    \includegraphics[width=0.9\columnwidth]{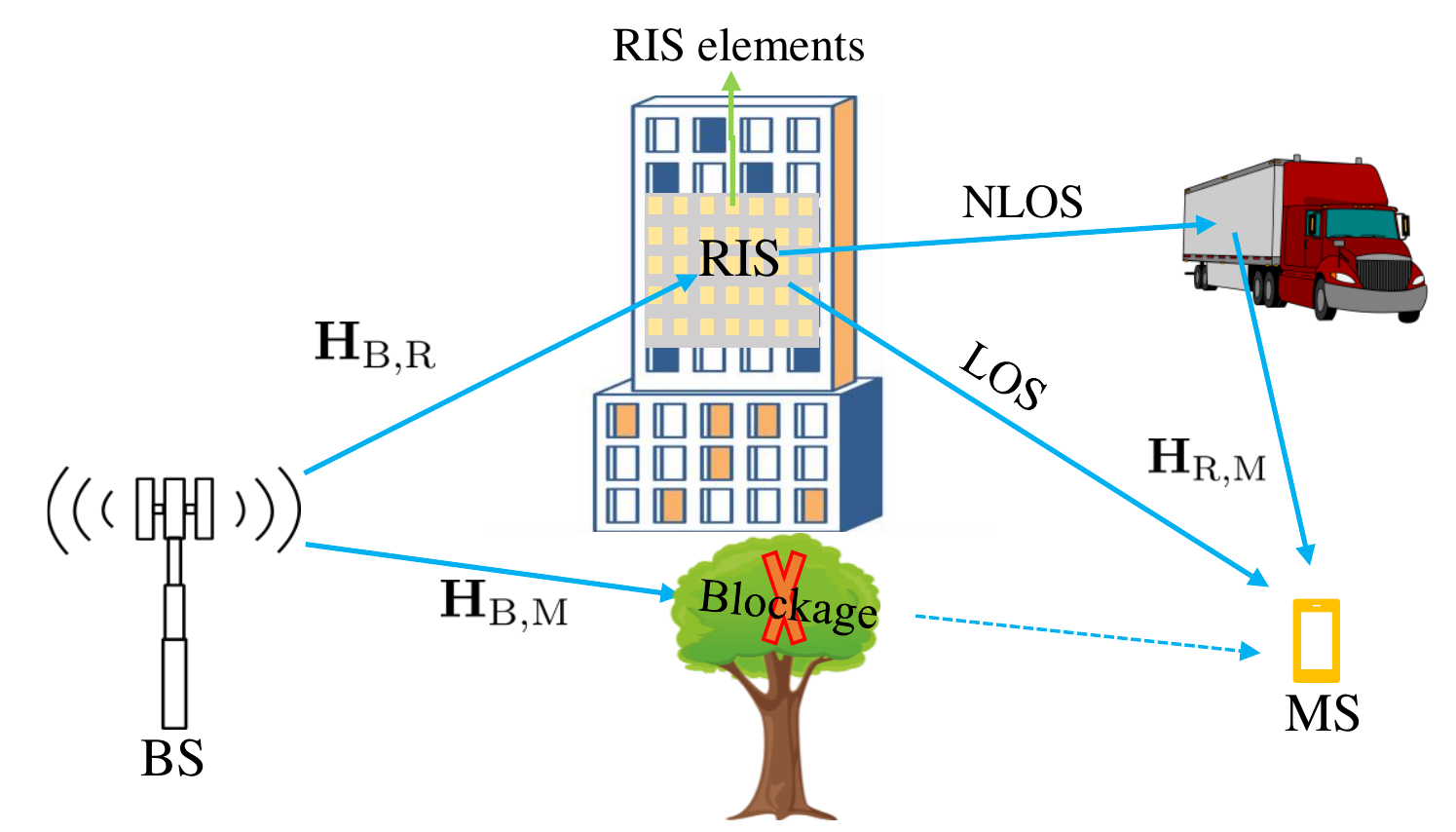}
    \caption{The RIS-assisted communication system in the presence of LOS blockage. Different types of RIS architectures can be considered.}
    \label{fig:system_model}
    \vspace{-0.6cm}
\end{figure}

In the sequel, the signal processing techniques are based on the system model, depicted in Fig.~\ref{fig:system_model}, where the BS communicates with the MS via the RIS. The number of (antenna) elements at the BS, MS, and RIS are denoted by $N_t$, $N_r$, and $N$, respectively. The RIS can be purely \emph{passive}, \emph{hybrid} (a combination of passive and active RIS elements), or in the form of \emph{hybrid relay-reflecting} (HR-RIS), i.e., a mixture functionality of a RIS and a relay~\cite{Yildirim2021,nguyen2021hybrid,nguyen2021spectral}. The antenna configuration at the BS and MS depends on the practical applications, e.g., carrier frequency, physical size constraint, and power consumption limitation, etc. In this paper, a multi-antenna BS is assumed while both single- and multi-antenna MSs are considered, i.e., $N_t > 1, N_r \geq 1$.

For the purely passive RIS architectures without any baseband processing units, CE can only be done at BS or MS due to the unavailability of observations at the RIS. According to the literature, the CE methods are classified into two major categories: individual CE (also known as cascaded CE)~\cite{He2020,Wei2020parallel, Wang-9103231,dearaujo2020parafacbased}, and channel parameter estimation~\cite{he2020channel,he2020channelANM,Ardah2020trice,schroeder2020passive}. For the former, parallel factor tensor decomposition, matrix factorization \& matrix completion and their variants can be applied. For the latter, inherent channel sparsity is adopted for the estimation algorithm design along with the assumption of parametric channel models. Compared to the individual CE, channel parameter estimation has the following advantages. \emph{(i)} Instead of estimating the individual channel matrices, the number of parameters to be estimated is reduced significantly.
\emph{(ii)} The inherent channel sparsity can be leveraged in the CE algorithm development, and the availability of various CS techniques can be applied or modified for extracting the channel parameters with a small amount of training overhead.  

On the contrary, for the ease of CE, active sensors can be added to the RIS, resulting in so-called hybrid RIS architecture~\cite{Alexandropoulos2020}. With the introduction of active sensors (elements), the CE can be performed at the RIS, and the CE problem degrades to that for P2P channels. However, this requires more computational power, hardware complexity, and energy consumption for the RIS. 

\subsection{Individual Channel Estimation}
In~\cite{Wang-9103231}, Wang \textit{et al.} studied CE for the individual channels by using compressive sensing techniques, e.g., OMP and generalized AMP (GAMP). However, all the angular parameters were assumed to lie in a predefined grid, which leads to grid mismatch problem and limit its practical applications. In~\cite{Wei2020parallel, dearaujo2020parafacbased}, parallel factor tensor decomposition are applied to estimate the individual channel matrices, e.g., BS-RIS and RIS-MS channels with an iterative refinement of the individual CE conducted by using bilinear alternating least squares (BALS), while in~\cite{He2020} the individual channels are estimated in a sequential way by an alternation of matrix factorization and matrix completion. In these works, heavy training overhead is demanded as to achieve satisfactory estimation performance. 

\subsection{Channel Parameter Estimation}
Channel parameter estimation was considered in~\cite{he2020channel,he2020channelANM,Ardah2020trice,schroeder2020passive}. In~\cite{schroeder2020passive}, we made an intensive comparison on the CE performance for both passive and hybrid RIS architectures in terms of mean squared error (MSE) of the channel parameters, RIS gain, and SE. The CE for the passive and hybrid RIS architectures is addressed via atomic norm minimization along with multiple signal classification (MUSIC) algorithm or its variants. As for the passive RIS architecture, we proposed a two-stage sounding and CE procedure, detailed in~\cite{he2020channelANM}. By assuming downlink training (also applied for uplink training), the BS sends pilots to the MS, followed by the first stage of CE for the extraction of the angle of arrivals (AoAs) at the MS and angle of departure (AoD) at BS. The estimates are used to guide the design of the beam training matrix in the second stage sounding, aiming at the estimation of the angle differences associated with the RIS and products of propagation path gains. The hybrid RIS architecture considers alternating uplink and downlink training. Namely, the BS-RIS and RIS-MS channels are estimated at the RIS with pilots sent from the BS and MS, respectively.

\begin{figure}
    \vspace{-0.5cm}
    \centering
    \includegraphics[width=0.8\columnwidth]{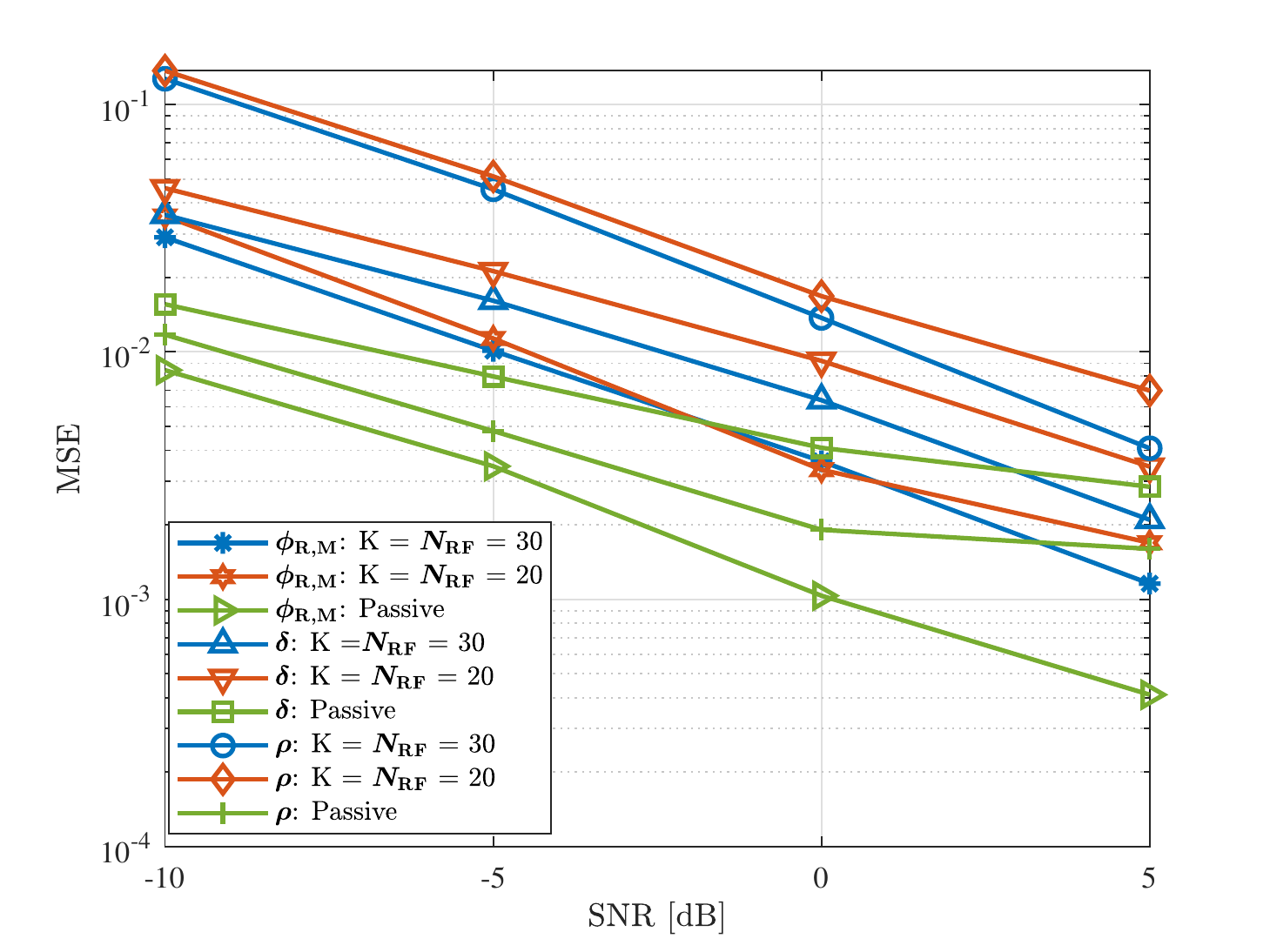}
    \caption{Passive RIS vs.\ Hybrid RIS regarding MSE of the channel parameters.}
    \label{fig:results_rafaela}
        \vspace{-0.6cm}
\end{figure}

Fig.~\ref{fig:results_rafaela} shows the simulation results for the estimation of the channel parameters, where $\boldsymbol{\phi}_{\text{R,M}}$, $\boldsymbol{\delta}$, and $\boldsymbol{\rho}$ denotes the AoAs at MS, angle differences, and products of path gains, respectively. The performance is evaluated by considering the purely passive RIS and the hybrid RIS ($K = N_{\text{RF}} = 30$ and $K = N_{\text{RF}} = 20$), where $K$ is the number of active elements, and $N_{\text{RF}}$ is the number of radio frequency (RF) chains at the RIS. The MSE of the AoA shows that the purely passive RIS achieves better performance than the hybrid RIS under the same training overhead. A similar conclusion can be drawn for the estimation of products of path gains.
However, the MSEs of the angle difference are approximate for the setup with $K = N_{\text{RF}}=30$ and passive RIS architectures. 

\subsection{Position-Aided Channel Estimation}
\begin{figure}[t!]
    \centering
    \belowcaptionskip = -15pt
    \includegraphics[width=0.70\linewidth ]{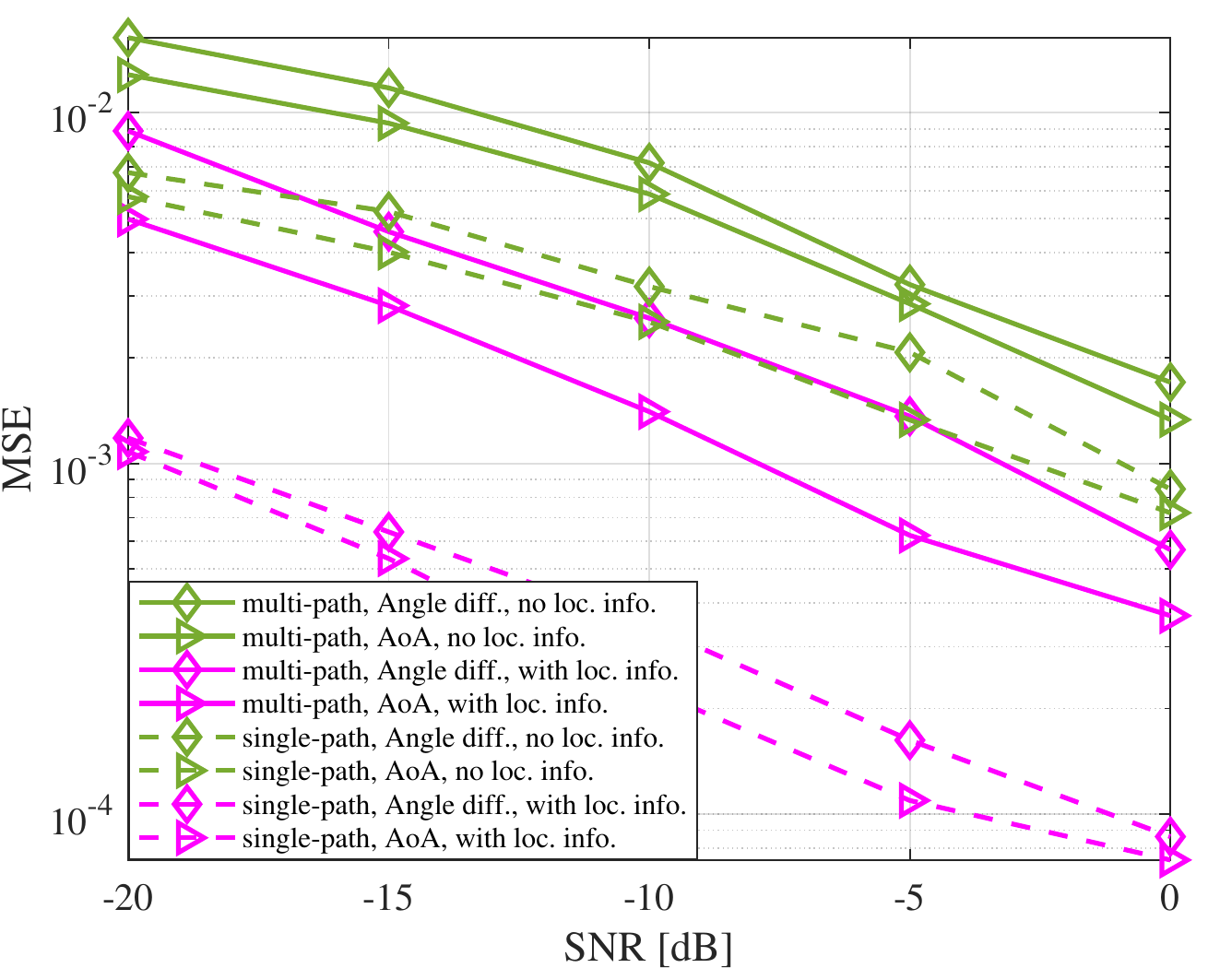}
    \caption{The effect of prior location information on CE performance, where $N_t = N_r = 16$ and $N = 64$~\cite{he2020leveraging}.}
    \label{fig:Effect_of_location_information}
\end{figure}

In the parametric channel models, the channel parameters are calculated based on the locations/coordinates of the terminals and vice versa. Thus, the position information can offer prior information, albeit rough, for the channel parameters, e.g., AoAs and AoDs. This rough information can enhance the design of training beams used for CE, which in turn brings better estimation performance~\cite{he2020leveraging,Garcia2016}. In~\cite{he2020leveraging}, we assumed a bounded position estimation error, which is equivalent to a reduced angular interval (compared to $[-\pi\;\pi]$) after mapping the position to angular information. 
It was verified that better parameter estimation performance can be attained. Relying on the parameter estimates, a better RIS phase control matrix can be designed for communication purpose. The effect of prior location information on CE performance is illustrated in Fig.~\ref{fig:Effect_of_location_information} for both single-path and multi-path scenarios. The training overhead in symbols are $T_t = 64$ and $T_t =144$, respectively. For both scenarios, obvious improvement can be seen for the MSE performance with the aid of prior location information.

\section{Hybrid Relay-Reflecting Intelligent Surface}

Intensive studies on the performance and design aspects of the RIS-assisted communication systems have shown the capability of RIS in improving the system capacity, sum-rate, and channel quality \cite{hu2017potential, hu2018beyond, jung2020asymptotic, ozdogan2020using, wu2019intelligent, zhang2020capacity}. However, a main limitation of the RIS is that its passive beamforming limits the beamforming gains. It has been shown in \cite{wu2019intelligent, bjornson_intelligent_2019} that a very large number of reflecting elements need to be deployed in the RIS to perform better than decode-and-forward (DF) relaying; otherwise, it can be easily outperformed even by a half-duplex (HD) relay with few elements. In addition, a practical RIS with limited-resolution phase shifts has considerable performance degradation \cite{guo2019weighted,zhang2020sum}. To overcome the aforementioned limitations of RIS, HR-RIS architectures and the concept of \emph{semi-active/passive beamforming} have been introduced in \cite{nguyen2021hybrid}. The basic idea of the HR-RIS is to connect a few elements of the RIS to RF chains and power amplifiers (PAs) (RF-PA chains). As a result, HR-RIS can leverage the advantages while mitigating the disadvantages of RIS and relay.

\begin{figure}[t]
    \belowcaptionskip = -15pt
	\centering
	\includegraphics[scale=0.5]{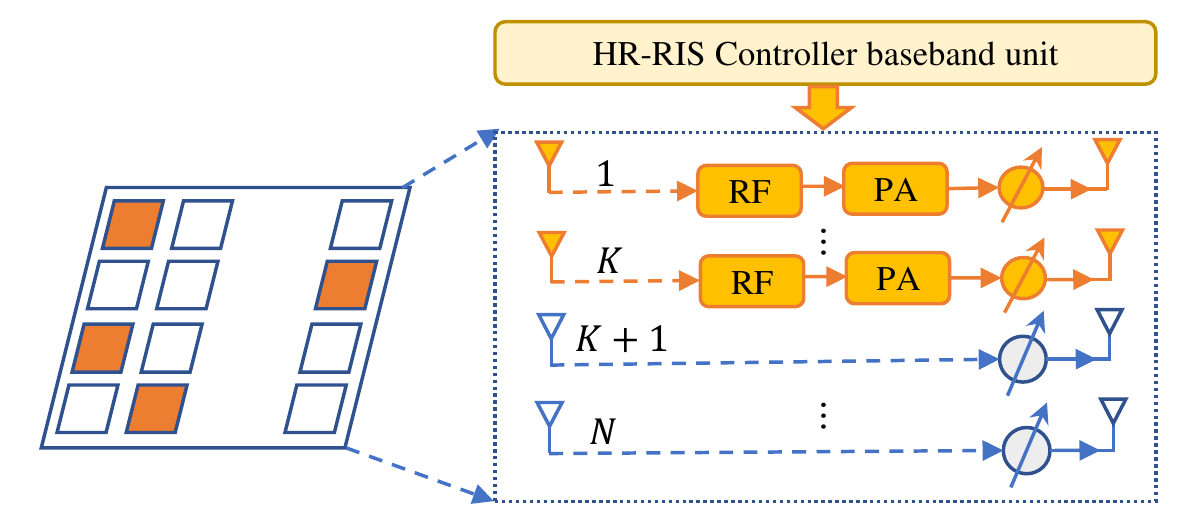}
	\caption{Illustration of the HR-RIS architecture. The dashed lines represents the dynamic connections in the dynamic HR-RIS architecture. In the fixed HR-RIS, they become fixed connections \cite{taha2019enabling, nguyen2021spectral}.}
	\label{fig_HRRIS_architecture}
\end{figure}

\subsection{Design Aspects of the HR-RIS}

With an HR-RIS equipped with $N$ elements, only $K \ll N$ active relay elements are deployed, while the remaining elements are passive and only reflect the incident signals. Similar to the conventional RIS, the passive element can only shift the phase; in contrast, the active one can tune both the phase and amplitude of the incident signal. By optimizing the number of active elements, the HR-RIS can achieve high active beamforming gains while still almost preserving the passive beamforming gain. The analysis and numerical results in \cite{nguyen2021hybrid} have shown that only a few active elements, i.e., small $K$, are sufficient for the HR-RIS to attain significant performance improvement with respect to the conventional RIS. In particular, increasing $K$ does not always guarantee the improvement in the SE, while causing a linear increase in the total power consumption and hardware cost. Considering these aspects, the HR-RIS is suggested to deploy with a single or few active elements to attain remarkable improvement in the SE and EE.

Two architectures for the HR-RIS, namely, the \emph{fixed} and \emph{dynamic} HR-RIS have been proposed in \cite{nguyen2021hybrid}. In the former, the number and the positions of the active elements are predetermined and designed in manufacture. By contrast, those in the dynamic HR-RIS can be changed according to the propagation condition. To adapt to the channel conditions, elements of the HR-RIS must be able to connect/disconnect from the RF-PA chains, which can be also turned on/off to save power. The fixed and dynamic HR-RIS architectures are illustrated in Fig.\ \ref{fig_HRRIS_architecture} \cite{taha2019deep, taha2019enabling}. 
\vspace{-0.1cm}
\subsection{Optimization of HR-RIS}

Unlike the conventional RIS, a subset of elements in the HR-RIS are active ones that are supposed to modify not only the phases but also the amplitudes of the incident signals. Therefore, their amplitudes are not unity and required to be optimized. In \cite{nguyen2021hybrid, nguyen2021spectral}, the optimization of the coefficients of HR-RIS is formulated in a SE maximization problem, which is solved with an alternating optimization (AO) method \cite{zhang2020capacity}. The closed-form solution to the reflecting/relaying coefficient of the fixed HR-RIS is derived. For the dynamic HR-RIS, the problem becomes more challenging because the set of active elements is unavailable. To overcome this, a power allocation problem is formulated, enabling an efficient selection of active elements and obtaining their amplitudes based on the well-known water filling algorithm \cite{zhang2020capacity}.

\begin{figure}[t]
    \belowcaptionskip = -15pt
    \centering
	\includegraphics[scale=0.45]{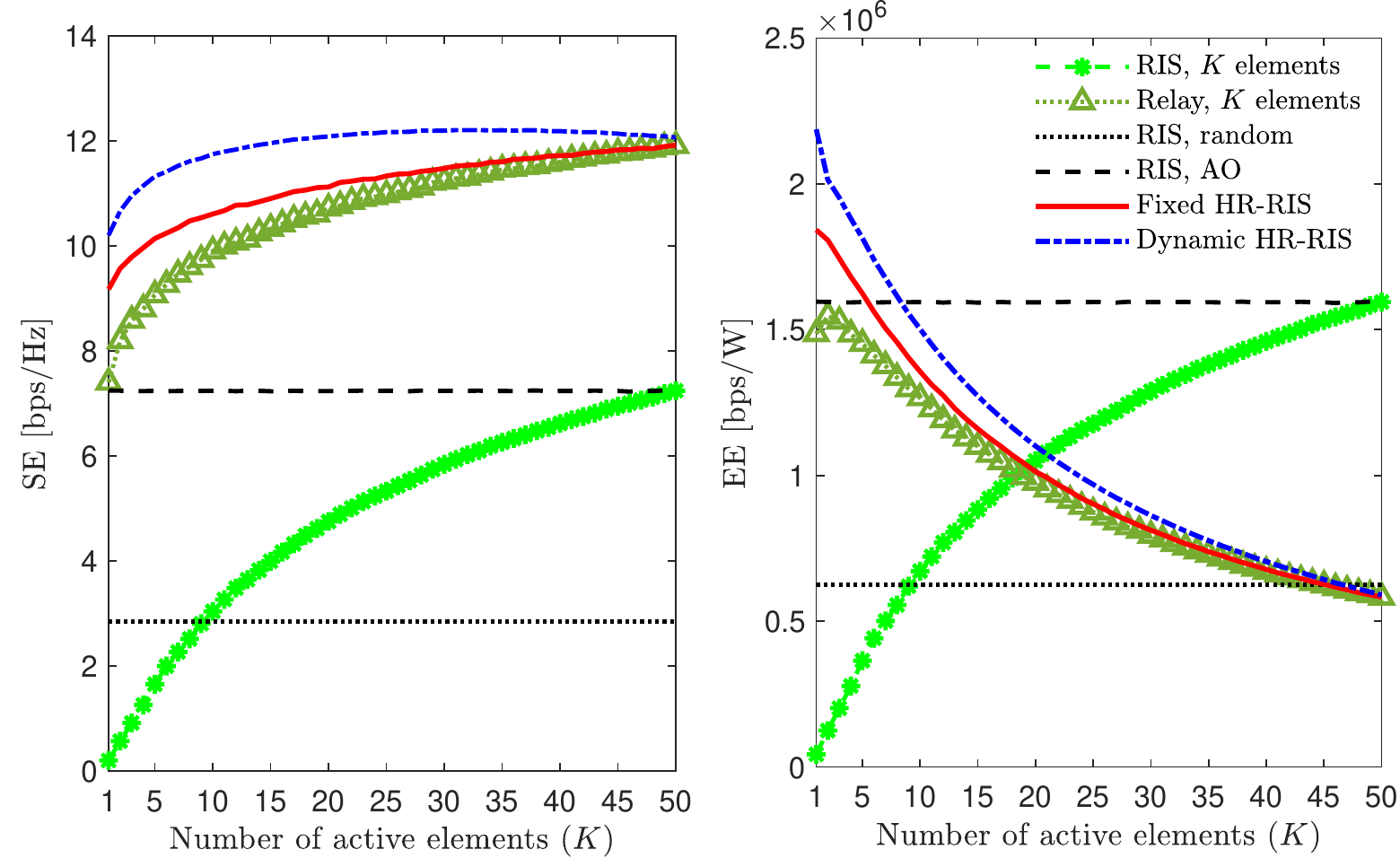}
	\caption{SE and EE improvement of the HR-RIS in a $4 \times 2$ MIMO system with $N = 50$. The transmit power of the BS and active elements at the HR-RIS are $30$ dBm and $0$ dBm, respectively \cite{nguyen2021hybrid}.}
	\label{fig_SE_EE_vs_K}
\end{figure}

The SE and EE of the proposed HR-RIS are shown Fig.\ \ref{fig_SE_EE_vs_K} in comparison with those of the conventional RIS, whose phases are either randomly generated or optimized using the AO scheme \cite{zhang2020capacity}. The results for the relay/RIS with only $K$, $K \in [1,N]$,  active/passive elements are also included for comparison. It is observed that increasing $K$ provides the RIS (with only $K$ elements) higher passive beamforming gains, and thus, its SE and EE  monotonically increases with $K$. By contrast, with a larger $K$, the relay and HR-RIS can significantly amplify the signal and attain higher SEs, which, however, comes at the cost of degraded EEs due to the increase in power consumption. Furthermore, it is observed that the SEs of the HR-RIS schemes quickly grow with small $K$, and almost converge with moderate or large $K$. In particular, the best SE-EE tradeoff is attained with a small $K$.

\section{Data-Driven Approaches}
Different from model-based approaches, data-driven schemes can be implemented with very low complexity (with only multiplication and addition operations, and element-wise activation) \cite{pham2020intelligent,nguyen2020deep,huang2019indoor,alexandropoulos2020phase}. The computation-heavy model training can be done offline. Data-driven approaches have been applied for CE and beamforming, RIS phase control, and  symbol detection, etc. For example, RIS CE with the application of DL framework was first studied in~\cite{Elbir-9090876}. To be specific, a twin convolutional neural network (CNN) is designed for mapping the received pilot signals to both direct and cascaded channels. The latest work on the application of deep neural network (DNN) to RIS phase control was included in~\cite{ozdogan2020deep}.



\begin{figure}[t!]
    \centering
    \belowcaptionskip = -5pt
    \includegraphics[scale=0.40]{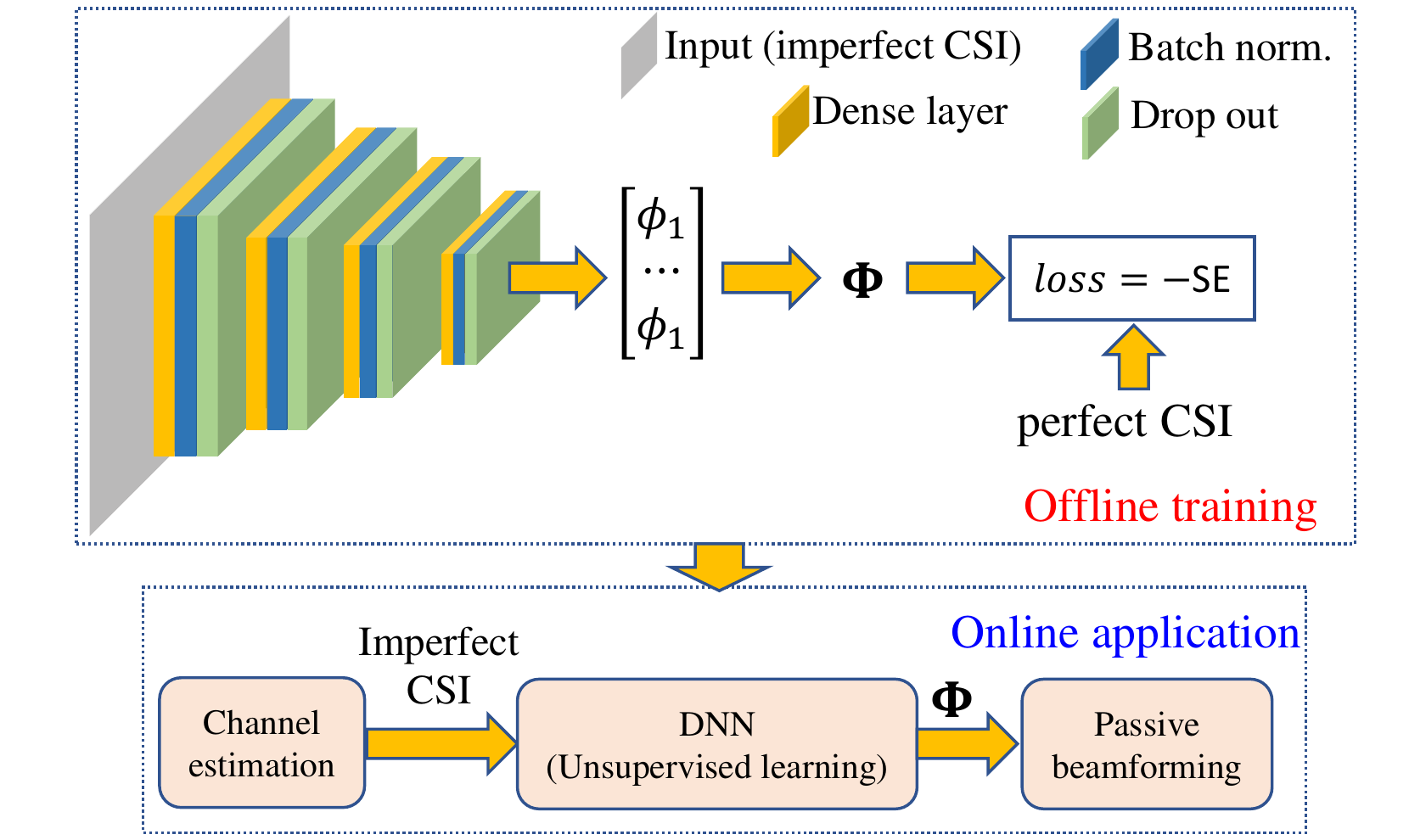}
    \caption{Illustration of the DNN-aided passive beamforming scheme.}
    \label{fig_DNN_model_illustration}
    \vspace{-0.4cm}
\end{figure}

\begin{figure}[t]
    \belowcaptionskip = -10pt
    \centering
	\includegraphics[scale=0.42]{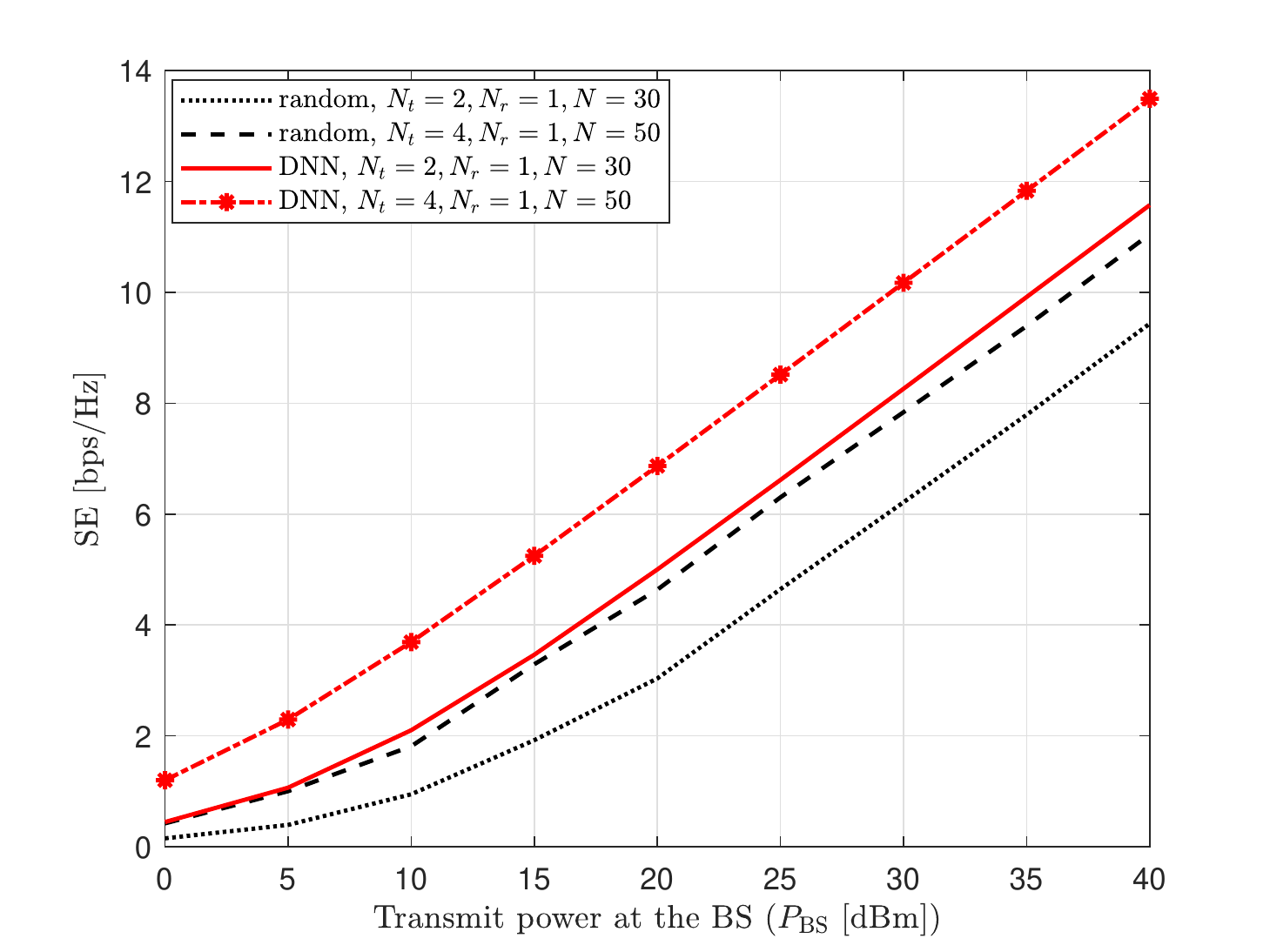}
	\caption{SE of the DNN-based passive beamforming method for $2 \times 1$ and $4 \times 1$ MISO systems with $N = \{30, 50\}$, respectively.}
	\label{fig_SE_DNN}
	 \vspace{-0.4cm}
\end{figure}

A DL model can also be trained for passive beamforming with imperfect CSI. It is noted that phase shifts  can be computed using closed-form solution obtained by traditionally hand-engineered methods, such as the semi-definite relaxation (SDR) \cite{wu2019intelligent} or AO \cite{zhang2020capacity}. However, these approaches generally require high complexity and feedback time, and thus, not amenable for real-time application \cite{gao2020unsupervised}. In particular, when the CSI errors exist, the conventional closed-form solutions derived for the perfect CSI may become unreliable. In this scenario, an unsupervised learning-based DNN-aided passive beamforming can be employed \cite{lin2019beamforming}. More specifically, the DNN is trained to learn the phase shifts with the imperfect CSI, as illustrated in Fig.\ \ref{fig_DNN_model_illustration}. In Fig.\ \ref{fig_SE_DNN}, we show the SEs of the $2 \times 1$ and $4 \times 1$ multiple-input single-output (MISO) systems assisted by the RISs with $N=30$ and $N=50$ elements, whose phase shifts are generated randomly or by DNN-aided passive beamforming method (Fig.\ \ref{fig_DNN_model_illustration}), with imperfect CSI. This scheme can achieve significant improvement in the SE with respect to the case of random phase shifts.

\section{Conclusion and Future Research Directions}
In this paper, we have covered a wide range of closely-related RIS topics with certain solutions given. Nevertheless, more efforts are required in order to make the commercialization of RIS feasible in the foreseeable future. For instance, mature yet valid RIS channel models for different frequency bands are still unavailable. For the moving users, efficient yet effective channel tracking algorithms need to be further investigated. Also, a full understanding of the pros and cons for different RIS architectures is not yet reached. The coexistence of multi-RIS with different architectures will make the communication systems even more complicated, joint RIS activation/deactivation control, phase profile design, and user scheduling also needs to be further studied.

\section*{Acknowledgment}
This work is supported by Horizon 2020, European Union's Framework Programme for Research and Innovation, under grant agreement no. 871464 (ARIADNE) and partially supported by the Academy of Finland 6Genesis Flagship (grant 318927) and by the Academy of Finland (ROHM project, grant 319485). We would like to thank Dr. Henk Wymeersch, the co-athor of \cite{he2020channelANM,he2020leveraging}, and Dr. Quang-Doanh Vu, Dr. Kyungchun Lee, the co-authors of \cite{nguyen2021hybrid, nguyen2021spectral}, for their contributions to the development of the CE schemes and the HR-RIS architectures introduced in this paper.


\ifCLASSOPTIONcaptionsoff
  \newpage
\fi



%
%
%

\bibliographystyle{IEEEtran}
\bibliography{IEEEabrv,Ref}

\begin{thebibliography}{10}
\providecommand{\url}[1]{#1}
\csname url@samestyle\endcsname
\providecommand{\newblock}{\relax}
\providecommand{\bibinfo}[2]{#2}
\providecommand{\BIBentrySTDinterwordspacing}{\spaceskip=0pt\relax}
\providecommand{\BIBentryALTinterwordstretchfactor}{4}
\providecommand{\BIBentryALTinterwordspacing}{\spaceskip=\fontdimen2\font plus
\BIBentryALTinterwordstretchfactor\fontdimen3\font minus
  \fontdimen4\font\relax}
\providecommand{\BIBforeignlanguage}[2]{{%
\expandafter\ifx\csname l@#1\endcsname\relax
\typeout{** WARNING: IEEEtran.bst: No hyphenation pattern has been}%
\typeout{** loaded for the language `#1'. Using the pattern for}%
\typeout{** the default language instead.}%
\else
\language=\csname l@#1\endcsname
\fi
#2}}
\providecommand{\BIBdecl}{\relax}
\BIBdecl

\bibitem{huang2019holographic}
C.~{Huang \textit{et al.}}, ``Holographic {MIMO} surfaces for {6G} wireless
  networks: Opportunities, challenges, and trends,'' \emph{{IEEE} Wireless
  Commun.}, vol.~27, no.~5, pp. 118--125, 2020.

\bibitem{Letaief-8808168}
K.~B. {Letaief}, W.~{Chen}, Y.~{Shi}, J.~{Zhang}, and Y.~A. {Zhang}, ``The
  roadmap to {6G: AI} empowered wireless networks,'' \emph{{IEEE} Commun.
  Mag.}, vol.~57, no.~8, pp. 84--90, 2019.

\bibitem{Wu-9326394}
Q.~{Wu}, S.~{Zhang}, B.~{Zheng}, C.~{You}, and R.~{Zhang}, ``Intelligent
  reflecting surface aided wireless communications: A tutorial,'' \emph{{IEEE}
  Trans. Commun.}, pp. 1--1, 2021.

\bibitem{Wu2020Mag}
Q.~{Wu} and R.~{Zhang}, ``Towards smart and reconfigurable environment:
  Intelligent reflecting surface aided wireless network,'' \emph{{IEEE} Commun.
  Mag.}, vol.~58, no.~1, pp. 106--112, 2020.

\bibitem{huang2018}
C.~{Huang}, A.~{Zappone}, G.~C. {Alexandropoulos}, M.~{Debbah}, and C.~{Yuen},
  ``Reconfigurable intelligent surfaces for energy efficiency in wireless
  communication,'' \emph{{IEEE} Trans. Wireless Commun.}, vol.~18, no.~8, pp.
  4157--4170, 2019.

\bibitem{Shlezinger2020}
N.~{Shlezinger}, G.~C. {Alexandropoulos}, M.~F. {Imani}, Y.~C. {Eldar}, and
  D.~R. {Smith}, ``Dynamic metasurface antennas for {6G} extreme massive {MIMO}
  communications,'' \emph{{IEEE} Wireless Commun.}, pp. 1--8, 2021.

\bibitem{wymeersch2019radio}
H.~{Wymeersch}, J.~{He}, B.~{Denis}, A.~{Clemente}, and M.~{Juntti}, ``Radio
  localization and mapping with reconfigurable intelligent surfaces:
  Challenges, opportunities, and research directions,'' \emph{{IEEE} Veh.
  Technol. Mag.}, vol.~15, no.~4, pp. 52--61, 2020.

\bibitem{He2019large}
J.~{He}, H.~{Wymeersch}, L.~{Kong}, O.~{Silv\'en}, and M.~{Juntti}, ``Large
  intelligent surface for positioning in millimeter wave {MIMO} systems,'' in
  \emph{proc. of IEEE VTC2020-Spring}, 2020, pp. 1--5.

\bibitem{he2019adaptive}
J.~{He}, H.~{Wymeersch}, T.~{Sanguanpuak}, O.~{Silv\'en}, and M.~{Juntti},
  ``Adaptive beamforming design for {mmWave RIS}-aided joint localization and
  communication,'' in \emph{proc. of IEEE WCNC Workshops (WCNCW)}, 2020, pp.
  1--6.

\bibitem{4286010-hum}
S.~V. {Hum}, M.~{Okoniewski}, and R.~J. {Davies}, ``Modeling and design of
  electronically tunable reflectarrays,'' \emph{{IEEE} Trans. Antennas
  Propag.}, vol.~55, no.~8, pp. 2200--2210, 2007.

\bibitem{Perez-Palomino-7109872}
G.~{Perez-Palomino \textit{et al.}}, ``Design and demonstration of an
  electronically scanned reflectarray antenna at 100 {GHz} using multiresonant
  cells based on liquid crystals,'' \emph{{IEEE} Trans. Antennas Propag.},
  vol.~63, no.~8, pp. 3722--3727, 2015.

\bibitem{Najafi2020}
M.~{Najafi}, V.~{Jamali}, R.~{Schober}, and H.~{Vincent Poor}, ``Physics-based
  modeling and scalable optimization of large intelligent reflecting
  surfaces,'' \emph{{IEEE} Trans. Commun.}, pp. 1--1, 2020.

\bibitem{Garcia2020}
J.~C.~B. {Garcia}, A.~{Sibille}, and M.~{Kamoun}, ``Reconfigurable intelligent
  surfaces: Bridging the gap between scattering and reflection,'' \emph{{IEEE}
  J. Sel. Areas Commun.}, vol.~38, no.~11, pp. 2538--2547, 2020.

\bibitem{ozdogan2020}
{\"O}.~{\"Ozdogan}, E.~{Bj\"ornson}, and E.~G. {Larsson}, ``Intelligent
  reflecting surfaces: Physics, propagation, and pathloss modeling,''
  \emph{{IEEE} Wireless Commun. Lett.}, vol.~9, no.~5, pp. 581--585, 2020.

\bibitem{Bjornson2021}
E.~{Bj\"ornson} and L.~{Sanguinetti}, ``Rayleigh fading modeling and channel
  hardening for reconfigurable intelligent surfaces,'' \emph{{IEEE} Wireless
  Commun. Lett.}, vol.~10, no.~4, pp. 830--834, 2021.

\bibitem{alexandropoulos2021reconfigurable}
G.~C. Alexandropoulos, N.~Shlezinger, and P.~del Hougne, ``Reconfigurable
  intelligent surfaces for rich scattering wireless communications: Recent
  experiments, challenges, and opportunities,'' \emph{arXiv}, 2021.

\bibitem{basar2020indoor}
E.~Basar and I.~Yildirim, ``Indoor and outdoor physical channel modeling and
  efficient positioning for reconfigurable intelligent surfaces in mm{W}ave
  bands,'' \emph{arXiv}, 2020.

\bibitem{Abeywickrama2020}
S.~{Abeywickrama}, R.~{Zhang}, Q.~{Wu}, and C.~{Yuen}, ``Intelligent reflecting
  surface: Practical phase shift model and beamforming optimization,''
  \emph{{IEEE} Trans. Commun.}, vol.~68, no.~9, pp. 5849--5863, 2020.

\bibitem{Wang-9103231}
P.~{Wang}, J.~{Fang}, H.~{Duan}, and H.~{Li}, ``Compressed channel estimation
  for intelligent reflecting surface-assisted millimeter wave systems,''
  \emph{IEEE Signal Processing Letters}, vol.~27, pp. 905--909, 2020.

\bibitem{he2020channel}
J.~{He}, M.~{Leinonen}, H.~{Wymeersch}, and M.~{Juntti}, ``Channel estimation
  for {RIS}-aided {mmWave MIMO} channels,'' in \emph{proc. IEEE Global
  Communications Conference}, 2020, pp. 1--6.

\bibitem{he2020channelANM}
J.~{He}, H.~{Wymeersch}, and M.~{Juntti}, ``Channel estimation for {RIS}-aided
  {mmWave MIMO} systems via atomic norm minimization,'' \emph{{IEEE} Trans.
  Wireless Commun.}, pp. 1--1, 2021.

\bibitem{Ardah2020trice}
K.~{Ardah}, S.~{Gherekhloo}, A.~L.~F. {de Almeida}, and M.~{Haardt}, ``{TRICE}:
  An efficient channel estimation framework for {RIS}-aided {MIMO}
  communications,'' \emph{{IEEE} Signal Process. Lett.}, vol.~28, pp. 513--517,
  2021.

\bibitem{schroeder2020passive}
R.~Schroeder, J.~He, and M.~Juntti, ``Passive {RIS} vs.hybrid {RIS}: A
  comparative study on channel estimation,'' \emph{arXiv}, 2020.

\bibitem{Wei2021}
L.~{Wei}, C.~{Huang}, G.~C. {Alexandropoulos}, C.~{Yuen}, Z.~{Zhang}, and
  M.~{Debbah}, ``Channel estimation for {RIS}-empowered multi-user {MISO}
  wireless communications,'' \emph{{IEEE} Trans. Commun.}, pp. 1--1, 2021.

\bibitem{he2020leveraging}
J.~{He}, H.~{Wymeersch}, and M.~{Juntti}, ``Leveraging location information for
  {RIS}-aided {mmWave MIMO} communications,'' \emph{{IEEE} Wireless Commun.
  Lett.}, pp. 1--1, 2021.

\bibitem{GLYBOVSKI20161}
S.~B. Glybovski, S.~A. Tretyakov, P.~A. Belov, Y.~S. Kivshar, and C.~R.
  Simovski, ``Metasurfaces: From microwaves to visible,'' \emph{Physics
  Reports}, vol. 634, pp. 1--72, 2016.

\bibitem{LIASKOS20191}
C.~Liaskos, S.~Nie, A.~Tsioliaridou, A.~Pitsillides, S.~Ioannidis, and
  I.~Akyildiz, ``A novel communication paradigm for high capacity and security
  via programmable indoor wireless environments in next generation wireless
  systems,'' \emph{Ad Hoc Networks}, vol.~87, pp. 1--16, 2019.

\bibitem{9140329-renzo}
M.~{Di Renzo \textit{et al.}}, ``Smart radio environments empowered by
  reconfigurable intelligent surfaces: How it works, state of research, and the
  road ahead,'' \emph{{IEEE} J. Sel. Areas Commun.}, vol.~38, no.~11, pp.
  2450--2525, 2020.

\bibitem{Akdeniz2014}
M.~R. {Akdeniz et al.}, ``{Millimeter Wave Channel Modeling and Cellular
  Capacity Evaluation},'' \emph{{IEEE J. Sel. Areas Commun.}}, vol.~32, no.~6,
  pp. 1164--1179, June 2014.

\bibitem{Haneda2016}
K.~{Haneda et al.}, ``{Indoor 5G 3GPP-like channel models for office and
  shopping mall environments},'' in \emph{{Proc. ICC Workshops}}, May 2016, pp.
  1--6.

\bibitem{Ntontin2020}
K.~Ntontin, A.-A.~A. Boulogeorgos, D.~Selimis, F.~Lazarakis, A.~Alexiou, and
  S.~Chatzinotas, ``Reconfigurable intelligent surface optimal placement in
  millimeter-wave networks,'' \emph{arXiv:2011.09949}, pp. 1--14, 2020.

\bibitem{Yildirim2021}
I.~{Yildirim}, F.~{Kilinc}, E.~{Basar}, and G.~C. {Alexandropoulos}, ``Hybrid
  {RIS}-empowered reflection and decode-and-forward relaying for coverage
  extension,'' \emph{{IEEE} Commun. Lett.}, pp. 1--1, 2021.

\bibitem{nguyen2021hybrid}
\BIBentryALTinterwordspacing
N.~T. Nguyen, Q.-D. Vu, K.~Lee, and M.~Juntti, ``{Hybrid relay-reflecting
  intelligent surface-assisted wireless communication},'' \emph{arXiv}, 2021.
  [Online]. Available: \url{https://arxiv.org/abs/2103.03900}
\BIBentrySTDinterwordspacing

\bibitem{nguyen2021spectral}
------, ``{Spectral efficiency optimization for hybrid relay-reflecting
  intelligent surface},'' \emph{IEEE Int. Conf. Commun. Workshops (ICCW)},
  2021.

\bibitem{He2020}
Z.~{He} and X.~{Yuan}, ``Cascaded channel estimation for large intelligent
  metasurface assisted massive {MIMO},'' \emph{{IEEE} Wireless Commun. Lett.},
  vol.~9, no.~2, pp. 210--214, Feb 2020.

\bibitem{Wei2020parallel}
L.~{Wei}, C.~{Huang}, G.~C. {Alexandropoulos}, and C.~{Yuen}, ``Parallel factor
  decomposition channel estimation in {RIS}-assisted multi-user {MISO}
  communication,'' in \emph{proc. IEEE 11th Sensor Array and Multichannel
  Signal Processing Workshop (SAM)}, 2020, pp. 1--5.

\bibitem{dearaujo2020parafacbased}
G.~T. {de Araújo} and A.~L.~F. {de Almeida}, ``{PARAFAC}-based channel
  estimation for intelligent reflective surface assisted {MIMO} system,'' in
  \emph{proc. IEEE 11th Sensor Array and Multichannel Signal Processing
  Workshop (SAM)}, 2020, pp. 1--5.

\bibitem{Alexandropoulos2020}
G.~C. {Alexandropoulos} and E.~{Vlachos}, ``A hardware architecture for
  reconfigurable intelligent surfaces with minimal active elements for explicit
  channel estimation,'' in \emph{proc. of IEEE International Conference on
  Acoustics, Speech and Signal Processing (ICASSP)}, 2020, pp. 9175--9179.

\bibitem{Garcia2016}
N.~{Garcia}, H.~{Wymeersch}, E.~G. {Str{\"o}m}, and D.~{Slock},
  ``Location-aided mm-wave channel estimation for vehicular communication,'' in
  \emph{proc. of IEEE International Workshop on Signal Processing Advances in
  Wireless Communications (SPAWC)}, 2016, pp. 1--5.

\bibitem{hu2017potential}
S.~Hu, F.~Rusek, and O.~Edfors, ``The potential of using large antenna arrays
  on intelligent surfaces,'' in \emph{IEEE 85th Veh. Tech. Conf. (VTC Spring)},
  2017, pp. 1--6.

\bibitem{hu2018beyond}
------, ``{Beyond massive MIMO: The potential of data transmission with large
  intelligent surfaces},'' \emph{{IEEE} Trans. Signal Process.}, vol.~66,
  no.~10, pp. 2746--2758, 2018.

\bibitem{jung2020asymptotic}
M.~Jung, W.~Saad, M.~Debbah, and C.~S. Hong, ``{Asymptotic optimality of
  reconfigurable intelligent surfaces: Passive beamforming and achievable
  rate},'' in \emph{IEEE Int. Conf. Commun. (ICC)}, 2020, pp. 1--6.

\bibitem{ozdogan2020using}
{\"O}.~{\"O}zdogan, E.~Bj{\"o}rnson, and E.~G. Larsson, ``{Using intelligent
  reflecting surfaces for rank improvement in MIMO communications},'' in
  \emph{IEEE Int. Conf. Acoustics, Speech and Signal Process. (ICASSP)}, 2020,
  pp. 9160--9164.

\bibitem{wu2019intelligent}
Q.~Wu and R.~Zhang, ``Intelligent reflecting surface enhanced wireless network
  via joint active and passive beamforming,'' \emph{{IEEE} Trans. Wireless
  Commun.}, vol.~18, no.~11, pp. 5394--5409, 2019.

\bibitem{zhang2020capacity}
S.~Zhang and R.~Zhang, ``Capacity characterization for intelligent reflecting
  surface aided {MIMO} communication,'' \emph{{IEEE} J. Sel. Areas Commun.},
  vol.~38, no.~8, pp. 1823--1838, 2020.

\bibitem{bjornson_intelligent_2019}
E.~Bj\"{o}rnson, O.~{\"O}zdogan, and E.~G. Larsson, ``Intelligent reflecting
  surface vs. decode-and-forward: How large surfaces are needed to beat
  relaying?'' \emph{IEEE Wireless Commun. Lett.}, pp. 1--1, 2019.

\bibitem{guo2019weighted}
H.~Guo, Y.-C. Liang, J.~Chen, and E.~G. Larsson, ``{Weighted Sum-Rate
  Maximization for Intelligent Reflecting Surface Enhanced Wireless
  Networks},'' in \emph{IEEE Global Commun. Conf. (GLOBECOM)}.\hskip 1em plus
  0.5em minus 0.4em\relax IEEE, 2019, pp. 1--6.

\bibitem{zhang2020sum}
Y.~Zhang, C.~Zhong, Z.~Zhang, and W.~Lu, ``{Sum rate optimization for two way
  communications with intelligent reflecting surface},'' \emph{IEEE Commun.
  Lett.}, vol.~24, no.~5, pp. 1090--1094, 2020.

\bibitem{taha2019enabling}
T.~Abdelrahman, A.~Muhammad, and A.~Ahmed, ``{Enabling large intelligent
  surfaces with compressive sensing and deep learning},'' \emph{arXiv preprint
  arXiv:1904.10136}, 2019.

\bibitem{taha2019deep}
------, ``{Deep learning for large intelligent surfaces in millimeter wave and
  massive MIMO systems},'' in \emph{IEEE Global Commun. Conf. (GLOBECOM)},
  2019, pp. 1--6.

\bibitem{pham2020intelligent}
Q.-V. Pham, N.~T. Nguyen, T.~Huynh-The, L.~B. Le, K.~Lee, and W.-J. Hwang,
  ``{Intelligent Radio Signal Processing: A Contemporary Survey},'' \emph{arXiv
  preprint arXiv:2008.08264}, 2020.

\bibitem{nguyen2020deep}
N.~T. Nguyen and K.~Lee, ``{Deep learning-aided tabu search detection for large
  MIMO systems},'' \emph{IEEE Trans. Wireless Commun.}, vol.~19, no.~6, pp.
  4262--4275, 2020.

\bibitem{huang2019indoor}
C.~Huang, G.~C. Alexandropoulos, C.~Yuen, and M.~Debbah, ``Indoor signal
  focusing with deep learning designed reconfigurable intelligent surfaces,''
  in \emph{IEEE 20th Int. Workshop Signal Process. Advances Wireless Commun.
  (SPAWC)}, 2019, pp. 1--5.

\bibitem{alexandropoulos2020phase}
G.~C. Alexandropoulos, S.~Samarakoon, M.~Bennis, and M.~Debbah, ``{Phase
  Configuration Learning in Wireless Networks with Multiple Reconfigurable
  Intelligent Surfaces},'' 2020.

\bibitem{Elbir-9090876}
A.~M. {Elbir}, A.~{Papazafeiropoulos}, P.~{Kourtessis}, and S.~{Chatzinotas},
  ``Deep channel learning for large intelligent surfaces aided mm-wave massive
  {MIMO} systems,'' \emph{{IEEE} Wireless Commun. Lett.}, vol.~9, no.~9, pp.
  1447--1451, 2020.

\bibitem{ozdogan2020deep}
{\"O}.~{\"Ozdogan} and E.~Bj\"ornson, ``Deep learning-based phase
  reconfiguration for intelligent reflecting surfaces,'' \emph{arXiv}, 2020.

\bibitem{gao2020unsupervised}
J.~Gao, C.~Zhong, X.~Chen, H.~Lin, and Z.~Zhang, ``Unsupervised learning for
  passive beamforming,'' \emph{IEEE Commun. Lett.}, vol.~24, no.~5, pp.
  1052--1056, 2020.

\bibitem{lin2019beamforming}
T.~Lin and Y.~Zhu, ``{Beamforming design for large-scale antenna arrays using
  deep learning},'' \emph{IEEE Wireless Commun. Lett.}, vol.~9, no.~1, pp.
  103--107, 2019.

\end{thebibliography}

\end{document}